# Large-scale on-chip integration of gate-voltage addressable hybrid superconductor-semiconductor quantum wells field effect nano-switch arrays

Kaveh Delfanazari[1,2,3] *, Jiahui Li[3], Peng Ma[3], Reuben K. Puddy[3], Teng Yi[3], Yusheng Xiong[1], Ian Farrer[4], Sachio Komori[5,6], Jason Robinson[5], David A. Ritchie[3], Michael J. Kelly[2,3], Hannah J. Joyce[2], and Charles G. Smith[3]

[1]*Electronics and Nanoscale Engineering Division, James Watt School of Engineering, University of Glasgow, Glasgow G12 8QQ, UK*

[2]*Electrical Engineering Division, Engineering Department, University of Cambridge, Cambridge CB3 0FA, UK*

[3]*Department of Physics, Cavendish Laboratory, University of Cambridge, Cambridge CB3 0HE, UK*

[4]*Department of Electronic and Electrical Engineering, University of Sheffield, Mappin Street, Sheffield, S1 3JD, UK*

[5]*Department of Materials Science & Metallurgy, University of Cambridge, Cambridge CB3 0FS, UK*

[6]*Department of Physics, Nagoya University, Furo-cho, Chikusa-ku, Nagoya 464-8602, Japan*

*Corresponding author:* kaveh.delfanazari@glasgow.ac.uk   Dated: 07062023

**Stable, reproducible, scalable, addressable, and controllable hybrid superconductor-semiconductor (S-Sm) junctions and switches are key circuit elements and building blocks of gate-based quantum processors. The electrostatic field effect produced by the split gate voltages facilitates the realisation of nano-switches that can control the conductance or current in the hybrid S-Sm circuits based on 2D semiconducting electron systems. Here, we experimentally demonstrate a novel realisation of large-scale scalable, and gate voltage controllable hybrid field effect quantum chips. Each chip contains arrays of split gate field effect hybrid junctions, that work as conductance switches, and are made from $In_{0.75}Ga_{0.25}As$ quantum wells integrated with Nb superconducting electronic circuits. Each hybrid junction in the chip can be controlled and addressed through its corresponding source-drain and two global split gate contact pads that allow switching between their (super)conducting and insulating states. We fabricate a total of 18 quantum chips with 144 field effect hybrid Nb- $In_{0.75}Ga_{0.25}As$ 2DEG-Nb quantum wires and investigate the electrical response, switching voltage (on/off) statistics, quantum yield, and reproducibility of several devices at cryogenic temperatures. The proposed integrated quantum device architecture allows control of individual junctions in a large array on a chip useful for the development of emerging cryogenic nanoelectronics circuits and systems for their potential applications in fault-tolerant quantum technologies.**

**Introduction:**

Split-gate field effect transistors (FETs) consisting of source, drain, and split-gate electrodes are vital in integrated circuits, especially in facilitating energy-efficient high-speed switching in quantum hardware [1,2]. Semiconductors such as InGaAs commonly form the channel of a FET and separate the source and drain metal electrodes. A gate dielectric electrically isolates the channel from the split gate electrodes. Therefore, the efficient operation of FETs depends on effective electrostatic coupling between the electric field caused by the split gate voltage and the semiconducting channel [3]. A FET function as a conductance switch and operates in three regions.

Replacing the source-drain metal electrodes in conventional FETs with superconducting materials, such as niobium (Nb), will realise efficient and low-power consumption cryogenic logic devices such as field effect controlled superconducting quantum point contacts (SQPCs) and gate-voltage controlled Josephson junctions (Josephson junction transistors or Josephson FETs) [4-7]. Gate-controlled hybrid electronic devices have recently received considerable interest in quantum technology due to their unique technical advantages, e.g., they can work as conductance or current switches in cryogenic quantum circuits [8-10] or can be used as an artificial material platform for investigation of Andreev reflection, induced unconventional and topological superconductivity [11-18]. Such devices can therefore be engineered and integrated with wafer-scale semiconductor chips to form the building blocks of robust quantum computing systems. For these purposes, (i) high yields, such as the reliability of hybrid junction and quantum device fabrication processes, (ii) reproducibility of quantum phenomena from junction to junction and chip to chip, and (iii) manufacturability of the complex hybrid quantum circuits must be systematically investigated. Such evaluation of hybrid junctions and coherent quantum circuits will inform us if they can be integrated into a scalable architecture for their use in real superconducting quantum hardware.

Here, for the first time, we report successful micro and nanofabrication of large-scale arrays of field-effect hybrid devices on an In$_{0.75}$Ga$_{0.25}$As/GaAs semiconducting heterostructure chip. Our design and fabrication technique bring several advances to the scalable hybrid superconducting-semiconducting quantum circuits for the realisation of cryogenic quantum hardware with complex structures: (i) Our approach in using In$_{0.75}$Ga$_{0.25}$As two-dimensional electron gases (2DEG), which is located 120 nm below the wafer surface, while is challenging for circuit fabrication, opens up new opportunities for the integration of hybrid quantum circuits with InGaAs based advanced semiconducting electronic integrated circuits (IC). Moreover, the ability to tune the indium composition permits the formation of highly transmissive metal–semiconductor interfaces. (ii) Our superconducting materials have not been sputtered in the same chamber as the semiconducting wafers making the quantum circuit fabrication accessible, and user friendly as in situ deposition of superconducting and semiconducting films are expensive and out of the capability and affordability of most research groups. (iii) Using Nb, as a type II superconductor with a larger superconducting gap and higher transition temperature compared to commonly used in-situ fabricated Al-based circuits, allows the operation of hybrid quantum circuits at relatively high magnetic fields and high frequency making them compatible which most emerging quantum circuit platforms based on superconducting microwave coplanar waveguide resonators and circuit quantum electrodynamic systems. In fact, making high-quality Nb-2DEG interfaces, especially in large arrays, are much more challenging than Al-based hybrid interfaces, because of the nature of Nb-based hybrid interfaces, which we successfully overcome here. Our advanced techniques used in this work will help open up a new route towards the development of Nb-based hybrid quantum circuits. (iv) For the first time, we fabricated a hybrid quantum IC on a large scale that is addressable and controllable with voltage signals, that is robust again noise compared to current biased circuits, in a compound material platform with a complex design and fabrication process compatible with the most advanced semiconducting ICs. (v) The interesting properties of $In_xGa_{1-x}As$ 2DEG, such as low electron effective mass, large g-factor, and strong Rashba spin–orbit coupling, make them a very attractive material platform for applications in electronics, spintronics, and photonics topological quantum computing.

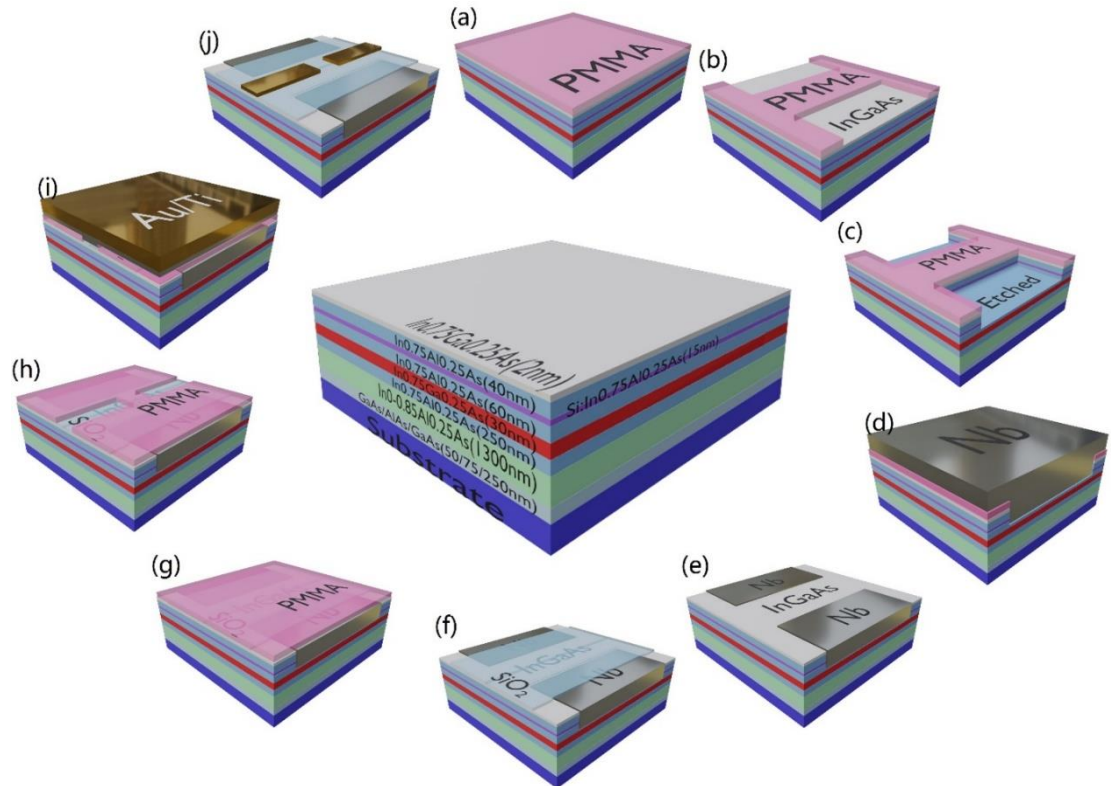

**Figure 1. Hybrid superconductor-semiconductor quantum chip fabrication process:** The schematic demonstration of a semiconducting wafer based on InGaAs/GaAs heterostructure is shown in the centre. The 2DEG is a 30 nm thick $In_{0.75}Ga_{0.25}As$ quantum well located 120nm beneath the surface. (a) PMMA spin-coated on the wafer. (b) Ebeam patterning is performed, and PMMA is developed to form the pattern. (c) Chemical wet etching to remove the layers above the 2DEG. (d) Nb is sputtered on the etched area of the wafer. (e) The Nb is lift-off, and the junction is formed. (f) A 50 nm thick $SiO_2$ oxide layer grown by chemical vapour deposition (CVD). (g) PMMA spin-coated. (h) Ebeam patterning and developing. (i) Au/Ti is thermally evaporated on the PMMA pattern. (j) The final step for a gated device fabrication after the lift-off of Au/Ti.

Therefore, the presented work is a considerable step towards realising advanced voltage-tunable, low-power cryogenic quantum hardware based on hybrid superconductor-semiconductor electronic circuits.

**Hybrid Quantum Chip Micro and Nanofabrication:**

The proposed quantum chip is based on $In_{0.75}Ga_{0.25}As$/GaAs heterostructure as shown in Fig. 1. The $In_{0.75}Ga_{0.25}As$ 2DEG with a 30nm thickness is buried 120 nm under the surface. The wafer is first chemically cleaned with Acetone, and IPA, respectively. Usually, an 30s of oxygen plasma ashing is included in the cleaning processes. The fabrication of split-gate hybrid junctions starts with the mesa section in which the 2DEG is isolated with respect to electrodes that are defined later. The PMMA is spin-coated on the wafer as shown in Fig. 1(a), and e-beam patterning is performed to define the superconducting electrode pads which are Nb in this work as shown in Fig 1(b). A chemical wet etching (see Fig. 1 (c)) followed by the formation of the 2DEG channel is performed before the Nb sputtering (Fig. 1 (d)), to form the hybrid Nb-2DEG-Nb junction as shown in Fig. 1 (e). The chemical wet etchant is $H_2SO_4:H_2O_2:H_2O= 1:8:1000$ with a typical etching rate of 1nm/s. Afterwards, a 50nm $SiO_x$ oxide layer grown via Chemical Vapour Deposition (CVD) is used to isolate the junction as shown as a transparent blue layer in Fig. 1 (f). Extra Buffered Hydrofluoric acid will open the window for electrodes. Then another PMMA e-beam pattern is performed to define the split gate as shown in Figs. 1 (e)-(h). Finally, the gate is defined via Au/Ti thermal evaporation followed by lift-off and a gated junction is fabricated as shown in Figs. 1 (i)-(j). The ohmic contact area is made of gold/germanium/nickel (AuGeNi) to form a low resistance and good chemical bond (adhesion) to the semiconductor substrate. These pads are placed 100 µm away from the junctions to reduce the impact of the normal electrons on the superconducting electrons. The ohmic is etched down to the 2DEG region to perform tunnelling spectroscopy measurements between Nb and 2DEG. For ease of demonstration, only the fabrication process of a single hybrid junction is shown in Fig. 1, however, the design allows wafer-scale nanofabrication.

A total of 144 gate-controllable hybrid S-Sm-S junctions are engineered and patterned on the surface of a semiconducting $In_{0.75}Ga_{0.25}As$/GaAs heterostructure wafer. Every 72 hybrid junctions are designed to fit across 9 smaller chips of $1 \times 1$ cm$^2$, each small chip containing an array of 8 hybrid double (16 single) interface junctions embedded in superconducting circuits.

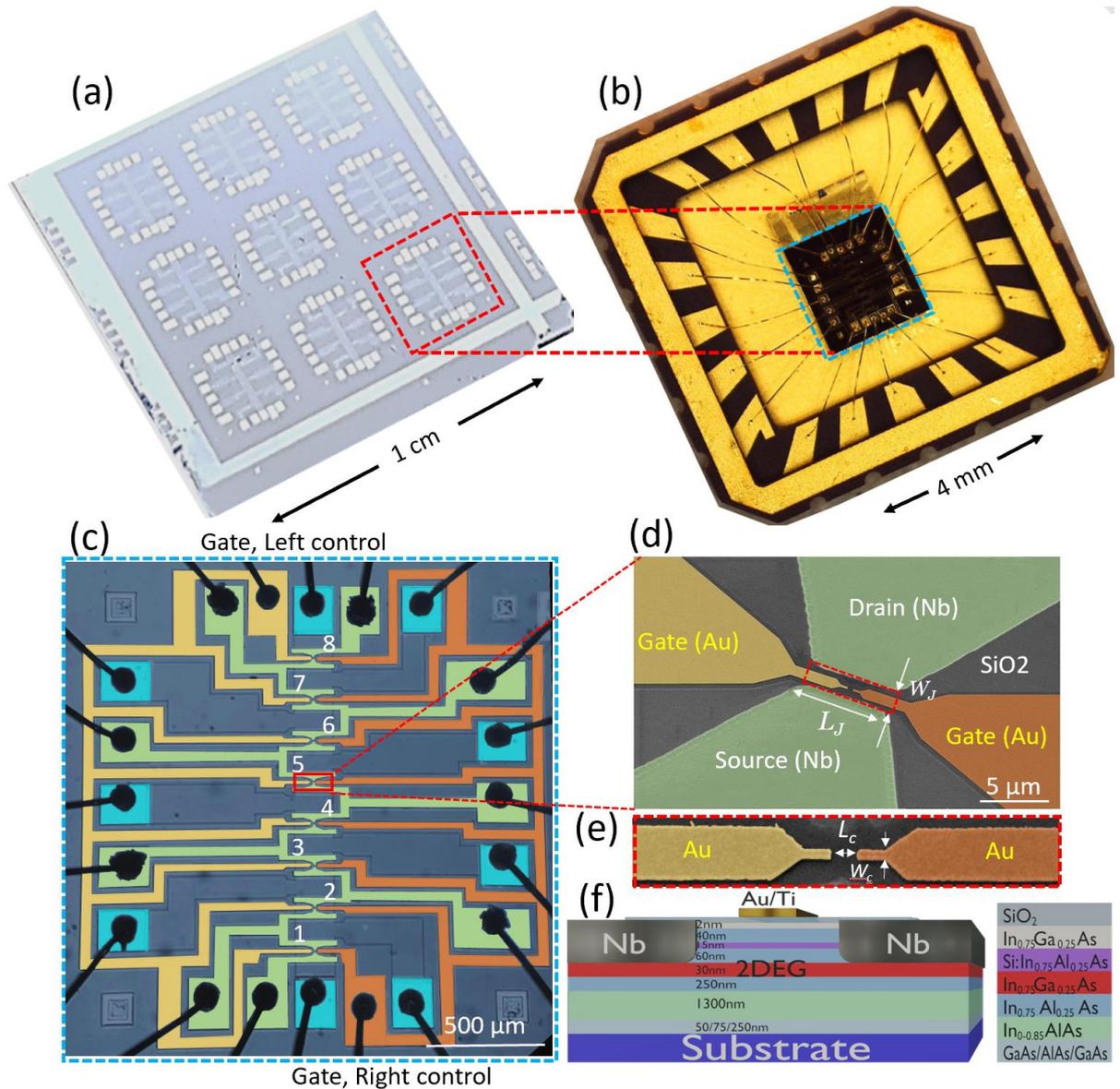

**Fig. 2 Gate voltage addressable hybrid quantum chip:** (a) The false-coloured optical image of a chip containing large-scale integration of hybrid semiconductor-superconductor junctions, before the final fabrication stages. The chip contains a total of 72 double interfaces (Nb-2DEG-Nb) and 144 single interfaces (Nb-2DEG or 2DEG-Nb) junctions. (b) A packaged addressable field effect quantum chip containing the chip shown in (c) after wirebonding to a leadless chip carrier (LCC). The chip is mounted in the cold finger of a cryostat or dilution refrigerator for the sub-Kelvin cryogenic test. (c) The false-coloured optical image of the addressable hybrid quantum chip containing an array of hybrid junctions made from superconducting Nb and high-quality $In_{0.75}Ga_{0.25}As$ two-dimensional electron gas (2DEG) in $In_{0.75}Al_{0.25}As$/GaAs heterostructure. Each junction in the chip can be addressed by applying voltages through two universal gate pads which are coloured yellow and orange. The ohmic contacts, shown with cyan pads, are etched down to the 2DEG region, for the purpose of tunnelling measurements between Nb and 2DEG (the data is not shown here). The area between Nb and ohmic pads is all etched away except the active area where hybrid junctions are formed (mesa area). (d) The false-coloured SEM image of one gate-voltage controlled hybrid symmetric and planar 2D junction on the quantum chip. (e) The dashed-red rectangle area in (d) is enlarged showing the split-gates area with a constriction width $W_c$ and length $L_c$ made from Au. A 50 nm thick $SiO_2$ dielectric was deposited to isolate the gates from the source-drain electrodes of the junctions. (f) Schematic illustration of a field effect hybrid switch from the side view with detailed semiconducting heterostructure nanolayers. The 30 nm $In_{0.75}Ga_{0.25}As$ 2DEG is shown in red.

The 8 hybrid junctions in each small chip are individually controlled by two global gate pads; one pad is designed to control the left side and the other pad to control the right side of each split gate in the superconducting integrated circuits. Some junctions failed during the fabrication or wire bonding process, and their data are excluded from this study. Figure 2 (a) shows the false-coloured optical image of a large-scale hybrid quantum chip before the final fabrication stages and dicing. The chip contains a total of 72 double interfaces (Nb-2DEG-Nb) junctions or 144 single interfaces (Nb-2DEG or 2DEG-Nb) junctions in a 1 × 1 cm$^2$ dimension. Figure 2 (b) shows a packaged addressable quantum chip containing the structure shown in Fig. 2 (c), which is wire-bonded to a leadless chip carrier (LCC) to be mounted in the cold finger of a cryostat or dilution refrigerator for characterisation and cryogenic measurements. A chip-integrated field effect addressable quantum IC with an array of 8 split gate hybrid double junctions is developed. It shows the false-coloured optical image of one of the nine completed devices shown in Fig. 2 (a) after dicing and wire-bonding. The hybrid junctions are made from superconducting Nb as their source and drain electrodes and high-quality In$_{0.75}$Ga$_{0.25}$As 2DEG as their active region [19].

The yellow and orange pads and lines in the chip (see Fig. 2 (c)) are the two universal gate pads in the chip addressing the junctions and controlling the Cooper-pairs density between the source and drain of each hybrid junction through the application of an external dc voltage across two sides of split gates (yellow for the left, and orange for the right split gate controllers, respectively). Figure 2 (d) is the false-coloured SEM image of one gate-voltage controlled field-effect hybrid symmetric and planar 2D junction on the chip. The dashed-red rectangle area in (d) is enlarged and shown in Fig. 2 (e), demonstrating the width $w_c$ and length $L_c$ of the constriction formed by the split gates. The side view of a field effect switch is schematically illustrated in Fig. 2 (f), where In$_{0.75}$Ga$_{0.25}$As 2DEG, Nb electrodes, and gates are shown in red, grey and yellow, respectively. The designed geometrical parameters of eight hybrid switches in each quantum chip are summarised in Table 1. The junction parameters are chosen to be large enough to investigate topological superconductivity in 2D systems [20-22] in the future generation of such devices. The geometrical parameters may vary from device to device after nanofabrication, especially for the length and width of the junctions, as the wet etch technique was used to take the unwanted semiconducting heterostructure areas and form source-drain leads and the active 2DEG region in between them. The fabricated junction parameters are commonly shorter than their designed dimensions [13-19].

**Table 1.** The designed geometrical parameters of eight hybrid field effect switches integrated into a single quantum chip.

| **Junctions** | $J_1$ | $J_2$ | $J_3$ | $J_4$ | $J_5$ | $J_6$ | $J_7$ | $J_8$ |
|---|---|---|---|---|---|---|---|---|
| $L_c$ (nm) | 400 | 400 | 400 | 400 | 400 | 400 | 400 | 400 |
| $W_c$ (nm) | 400 | 300 | 200 | 100 | 100 | 100 | 100 | 100 |
| $L_J$ (μm) | 1.4 | 1.4 | 1.4 | 1.4 | 1.4 | 1.4 | 1.4 | 3.2 |
| $W_J$ (μm) | 5 | 5 | 5 | 5 | 5 | 5 | 5 | 5 |

A depletion layer will form around the gate electrodes by applying a negative gate voltage to the split gates which defines a constriction in the quantum wells between two superconducting leads [23,24]. The constriction works as an electron waveguide (conducting quantum channel) in such a 2D electron system in our chips [25]. The split gate structure allows modulation of carrier density and mobility of the quantum wells beneath the gate electrodes and in the constriction area of the hybrid junction through the application of gate voltages. When the gate voltage is swept to negative values, the constriction length $L_c$ will reduce from its initial value $L_c$= 400 nm to $L_c$= 0 in perfect conditions. Therefore, the split gate is a knob to control the hybrid switches' electronic response. The carrier density and mobility of the In$_{0.75}$Ga$_{0.25}$As 2DEG in semiconducting heterostructure are defined by Shubnikov-de Haas measurement (see method section). The electron mass of the 2DEG is determined to be $m^*$= 0.039 $m_e$, where $m_e$ is the free electron mass [26]. From In$_{0.75}$Ga$_{0.25}$As 2DEG properties, the coherence length $\xi_N = \hbar v_F/2\pi k_B T \cong 200$ nm, and mean free path $l_e = e^{-1}\hbar\mu_e (2\pi n_s)^{1/2} \cong 2$ μm are calculated. Here, $v_F$ is the Fermi velocity. Consequently, the hybrid junctions fit into the clean limit regime ($l_e > \xi_N$) with ballistic transports ($l_e \gg L$) [13].

Figure 3 plots the switching response of eight field effect devices (J1-J8) fabricated in a single chip, with the dimensions given in Table 1. All junctions are measured twice: (i) the split gate voltage $V_g$ is swept from $V_g = 0$ to -1V (shown in black solid lines), and (ii) $V_g$ is returned from $V_g = -1V$ to 0 (shown in red solid lines). This measurement aims (i) to study the reproducibility of the conductance switching of each junction for two different sweep directions. (ii) to investigate the reproducibility of conductance switching in junctions of various dimensions (but similar geometry) in a single chip.

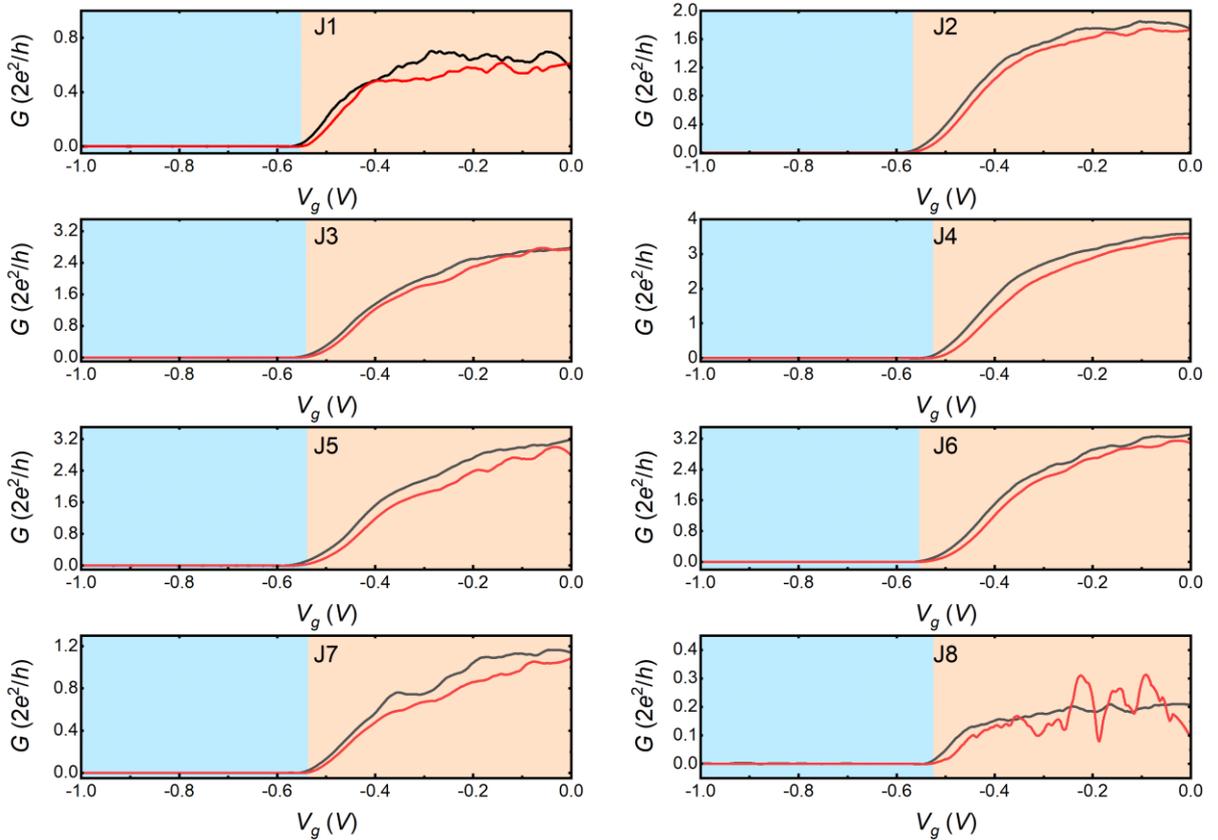

**Figure 3.** The conductance as a function of the electric field on the gates for eight (all) junctions in a single chip, measured at $T = 4K$. (a) J1, (b) J2, (c) J3, (d) J4, (e) J5, (f) J6, (g) J7, and (h) J8. There are two sweep directions from 0 to -1 V (black) and then from -1 V to 0 (red). Relatively good reproducibility of both effects are observed for all junctions in this chip.

We observe a relatively good reproducibility for two split gate voltage sweep directions (*L* and *R*) for all junctions in this and almost all measured chips. We observe conductance switching with very small hysteresis except for J8 which was purposely designed and fabricated larger than other junctions. This suggests that the quality and uniformity of the hybrid circuit gates, gate pads, and CVD-grown $SiO_2$ dielectric layer under gates used in the chips are of high fabrication qualities. Moreover, the gate voltage leakage and Joule heating in most junctions are negligible or small, so the conductance curves for both sweep directions almost overlap. Since the distance between the Nb leads ($L_J$= 3.2 μm) is long in J8, hot electrons seem to lead to Joule heating and hysteresis in the junction so the left and right conductance sweeps do not fully overlap, nevertheless, the switching voltage is not affected. The orange and blue areas in Fig. 3 show the hybrid switch ON and OFF states, respectively. Only a small dc switching voltage of $V_g$= -0.55 V is required to change the hybrid junction from conducting to isolating state. The ON state conductance $G_{ON}$ of the junctions varied by the junction's parameters, suggesting that the junctions' dimensions are altered during the fabrication process.

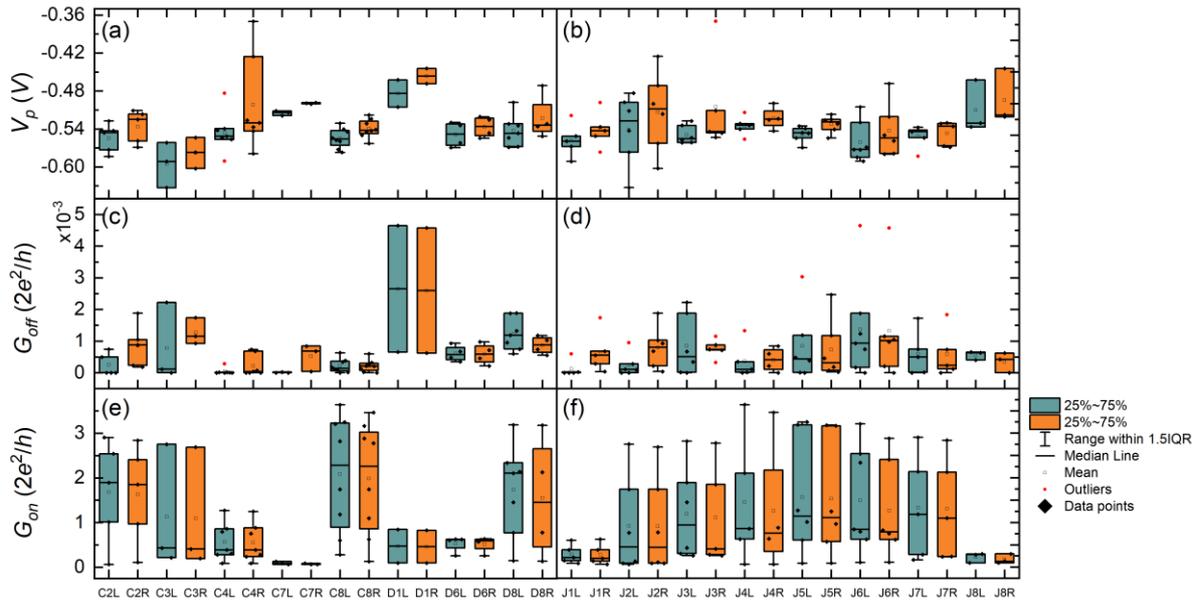

**Figure 4.** Switching (pinch-off) voltage of all measured chips (a), each with eight junctions (b) for different sweeping directions measured at $T$= 4K. Green colour boxes are the data for left sweeping gate voltage (0 to -1 V) and the orange boxes are the data for right sweeping gate voltage (-1 V to 0). (c) On-state conductance $G_{on}$ for all the measured chips, shown as C or D, each with eight junctions (d) measured at $T$= 4K. On-state refers to the height of the conductance at zero gate voltage. (e) Off-state conductance $G_{off}$ for all the measured chips, each with eight junctions (f) measured at $T$= 4K.

In Fig. 4, the green colour boxes are the data for left (0 to -1 V), and the orange boxes are the data for right (-1 V to 0) gate voltage sweep directions, respectively, with 25-75% data distribution in the boxes. 1.5IQR is that 95% of the data points are distributed in the range. The Median is the middle value of the data distribution. The mean is the average of the data points. Outliers are the data points outside the 1.5IQR. The pinch-off (switching) voltages $V_p$, and ON ($G_{ON}$) and OFF ($G_{OFF}$) state conductance values are studied for a large number of chips, and junctions as summarised in Fig. 4.

As shown in Figs. 4 (a) and (b), the average pinch-off voltage $V_p$ for both $L$ and $R$ sweep directions is around -0.56 V for more than 75% of all measured junctions in a large number of chips (note that some junctions are not measured due to failure reasons such as wire-bonding issues or broken contact pads). We observe relatively reproducible and consistent conductance switching response to the electric field induced by the split gates for several measured devices. This suggests that the next-generation hybrid nano-switches with improved design and performance could be used to control the cryogenic electronic hardware by the application of only -0.56 V or fewer dc voltages.

We also investigated the OFF-sate conductance $G_{OFF}$ of the hybrid junctions in measured chips and plotted them in Fig. 4 (c), and (d). The OFF-state conductance values for the majority of devices are almost zero soon after the application of ~ -0.56 V confirming the uniformity of large scale CVD-grown $SiO_2$ dielectrics on the wafer before dicing and forming the packaged quantum chips. Figure 4 suggests that the majority of the fabricated and measured field effect devices work as an ideal low dissipation switch under the application of split gate electric field in both voltage sweep directions. To analyse the manufacturability of the hybrid junctions we plot the ON state conductance $G_{ON}$ for a number of chips as shown in Fig. 4 (e), and (f). We find that $G_{ON}$ varies from chip to chip suggesting that the geometrical and interfacial parameters of the hybrid junctions may change during the fabrication process. In order to have a more uniform response attention should be paid to etch process optimisation to improve the hybrid junction interfaces. To be used in quantum circuits and processors, the large-scale manufacturability of hybrid conductance switches depends strongly on their reproducible and identical operational parameters. Here, we further investigate the reproducibility of switching voltage and ON-OFF state conductance parameters for several hybrid devices, both from the same and different chips, under multiple voltage sweeps (more than three times) as shown in Fig. 5. We find that despite slight variation due to fabrication errors, the devices show many repeatable behaviours.

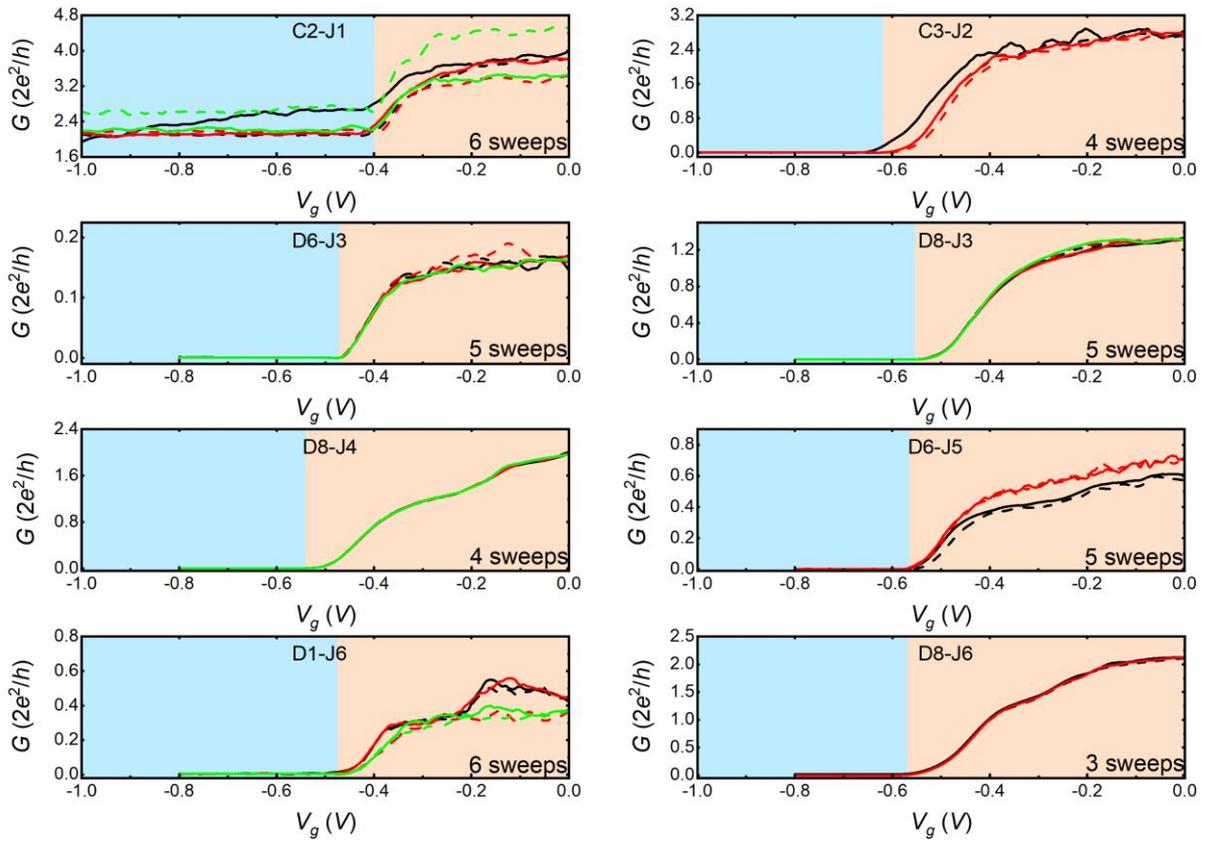

**Figure 5.** The conductance as a function of the electric field on the gates for devices with multiple sweeps to examine the reproducibility of the switching voltage, ON and OFF states conductance.

For example, although J1 of chip 2 does not completely pitch off even by the application of a large negative voltage, the conductance traces for six sweeps show slight deviation which could come from Joule heating due to the application of large voltages. The conductance switching failure of this device may be attributed to the formation of a different thickness of $SiO_2$ oxide layer compared to that of other junctions that show very good reproducibility and identical switching parameters (for instance see D8- J3, and D8- J4 & J6 for several sweeps). This non-uniformity may be due to forming of a thicker oxide layer, which prevents the hybrid junction from being completely pinched off and switching off from a conductive to an insulator state. We further investigate the relationship between the dimensions of gated hybrid junctions and their pinch-off voltages and show the results in Fig. 6. Specifically, the junction length $L_J$ is found to have a slight positive correlation with the pinch-off voltage, with longer junctions requiring a lower voltage to pinch off the channel.

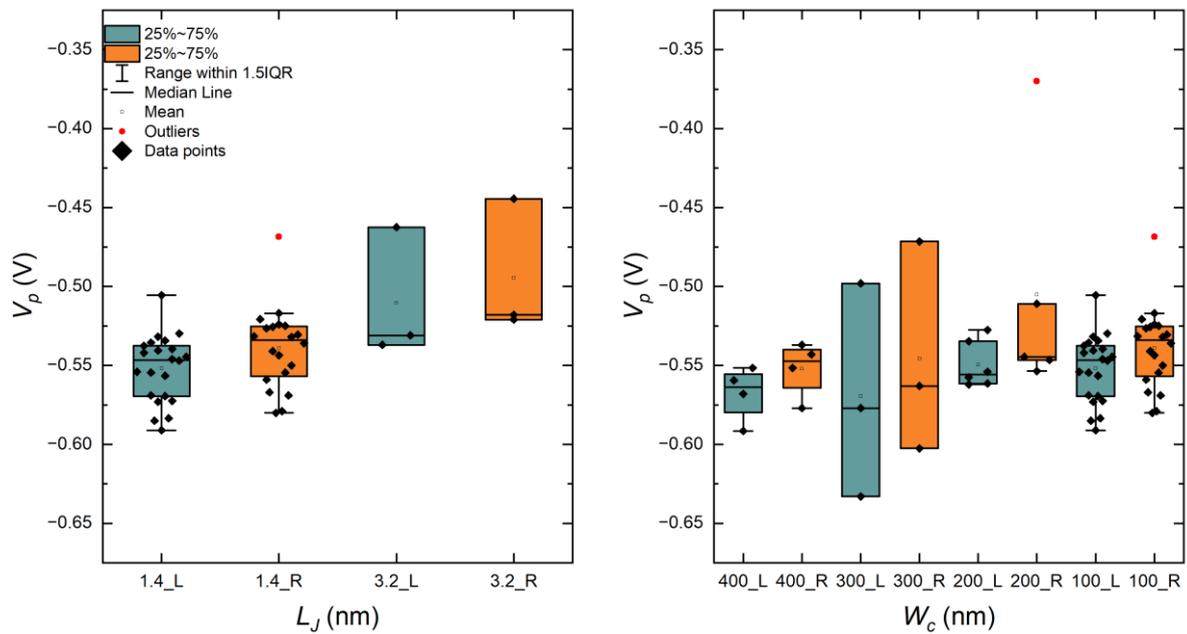

**Figure 6.** Pinch-off voltage $V_p$ as a function of junction length $L_J$. Green colour is the data for left sweeping gate voltage (0 to -1V), and the orange boxes are the data for right sweeping gate voltage (-1 to 0V). The data for chip 5 and chip 6 is absent. 1.5IQR is that 95% of the data points are distributed in the range. 25%-75% means 25-75% of data are in the box. The median is the middle value of the data distribution. Mean is the average of the data points. Outliers are the data points outside the 1.5IQR. (b) Pinch-off (switching) voltage $V_p$ as a function of constriction width $W_c$.

As shown in Fig. 6 (a), this positive correlation may be due to weaker proximity effects in the middle of the longer junctions under the gate, resulting from the extended junction length. Consequently, less gate perturbation may disturb the induced superconductivity, leading to an earlier pinch off of the channel. However, more data is needed to draw definitive conclusions about this relationship. In contrast, no obvious correlation is observed for constriction width

$W_c$. In theory, the width determines the length of the quasi-1D/1D channel, and if this length is much smaller than the scattering length/mean free path, the channel remains in the ballistic regime. Further theoretical analysis and simulations are needed to confirm this hypothesis. This will also motivate us to study these devices at lower temperatures to properly form the 1D channels. As these studies are not the focus of the present work will be discussed elsewhere. In Table 2, we summarise the yield with respect to the junction and constriction dimensions. The constriction width $W_c$ varies from 400 nm to 100 nm with fixed junction length $L_J$= 1.4 µm. As another control knob, junctions with $W_c$= 100 nm and $L_J$= 3.2 µm are fabricated to study the effects of the longer junctions. Such devices are found to show the minimum ON state conductance as discussed above and therefore offer the lowest yield. A slight reduction in quantum yield is observed with decreasing constriction width $W_c$.

**Table 2.** Junction length $L_J$ and constriction width $W_c$ correlated quantum yield. The yield is subdivided into different junction lengths and constriction sizes. The table sorts total switching devices versus measured devices with the specific size.

| $L_J$ (µm) | 1.4 | 1.4 | 1.4 | 1.4 | 3.2 |
|---|---|---|---|---|---|
| $W_c$ (nm) | 400 | 300 | 200 | 100 | 100 |
| Switching Devices | 6 | 6 | 6 | 26 | 5 |
| Measured Devices | 7 | 7 | 8 | 35 | 9 |
| Yield (%) | 85.7 | 85.7 | 75 | 74.3 | 55.6 |

Most of the non-switching junction with small $W_c$ shows negligible conductance at zero gate bias, which may be due to the surface charge accumulation under the split gates that are enough to pinch off the channel without extra gate bias or due to the electron leak into the gate rather than transport from source to drain. We observe a total yield of 74.24%, which counts for the working switches divided by the total number of measured devices, for a large array of hybrid devices as shown in Table 3. This includes yield for reproducibility of the conductance of individual hybrid switches when their split gates voltage is swept from 0 to -1V and then reversed from -1V to 0.

**Table 3.** Two sets of measured devices are labelled with 'C' and 'D'. The quantum yield counts for the working hybrid switch devices (showing switching response when the split gate voltage is swept to negative values) are divided by the total number of measured hybrid devices.

| Chip ID | C1 | C2 | C3 | C4 | C7 | C8 | D1 | D6 | D8 | D9 | Total |
|---|---|---|---|---|---|---|---|---|---|---|---|
| **Switching Devices** | 0 | 6 | 3 | 7 | 6 | 8 | 2 | 6 | 7 | 4 | 49 |
| **Measured Devices** | 8 | 7 | 8 | 8 | 8 | 8 | 2 | 6 | 7 | 4 | 66 |
| **Yield (%)** | 0 | 86 | 38 | 88 | 75 | 100 | 100 | 100 | 100 | 100 | 74.24 |

**Conclusion:**

We reported the first successful micro and nanofabrication of a large array of chip-integrated hybrid field effect quantum nanoelectronics devices and demonstrated a systematic experimental investigation of their conductance switching performance under the application of gate electric fields. We demonstrated techniques for the successful fabrication of novel cryogenic gate voltage addressable nanoelectronics chips with negligible gate voltage leakage and with high switching response statistics, reproducibility rate, and quantum yields. We found that to make efficient cryogenic switches, the attention should especially be on the quality junction geometrical and interfacial parameters as the former influence the uniform switching voltages and the latter have a direct effect on the ON-OFF state conductance. The OFF state conductance is also a function of the quality oxide layers isolating the source-drain electrodes of hybrid junctions from split gate electrodes. The techniques and experimental data presented here show that our field effect nano-switch devices, with modified designs and fabrication strategy, may help the development of novel cryogenic electronic switches for various classical or quantum cryogenic applications.

**Methods:**

1. **Device fabrication:**

The devices are based on MBE-grown In$_{0.75}$Ga$_{0.25}$As quantum wells in In$_{0.75}$Al$_{0.25}$As/GaAs high mobility semiconducting heterostructure. The wafer's detailed information has been discussed in our earlier works [26]. A three-inch wafer was diced into smaller pieces of 1.5 cm × 1.5 cm chips. From the bottom to the surface of the chip, it contains a GaAs substrate (500 µm), buffer layers of GaAs, AlAs, and GaAs (50 nm, 75 nm, and 250 nm), an InAlAs step-graded buffer layer (1300 nm), and an InAlAs buffer layer (250 nm). The 2DEG is a 30 nm thick In$_{0.75}$Ga$_{0.25}$As quantum well with electron density $n_s = 2.24 \times 10^{11}$ (cm$^{-2}$) and mobility $\mu_e = 2.5 \times 10^5$ (cm$^2$ V$^{-1}$ s$^{-1}$) in the dark and $n_s = 2.28 \times 10^{11}$ (cm$^{-2}$) and $\mu_e = 2.58 \times 10^5$ (cm$^2$ V$^{-1}$ s$^{-1}$) after light illumination. On top of the 2DEG quantum well, there are layers of In$_{0.75}$Al$_{0.25}$As spacer (60 nm), n-type modulation-doped In$_{0.75}$Al$_{0.25}$As (15 nm), In$_{0.75}$Al$_{0.25}$As layer (45 nm), and InGaAs cap layer (2 nm).

**Device characterisation and cryogenic measurements:**

The devices were characterised to ensure the height of resists, active regions, insulator thickness, etc, by Dektak Surface Profilometer, Veeco, before dicing. Each LCC leaded quantum chip was loaded into a 4K dip station for cryogenic tests. A standard lock-in technique was used to measure the quantum transport by superimposing a small ac-signal at a frequency of 70 Hz and an amplitude of 5 µV to the hybrid junction dc bias voltage and measuring the ac-current. The gates were swept by DC voltages supplied from NiDAQ national instruments modules that can provide low noise signals between ± 10 $V_{DC}$.